\documentstyle[preprint,aps]{revtex}
\tighten
\begin{document}

\title{Geometric Bounds in  
       Spherically Symmetric General Relativity}
\author{Jemal Guven${}^{(1)}$\thanks{e-mail \tt{ jemal@nuclecu.unam.mx}}
 and
   Niall \'O Murchadha${}^{(2)}$\thanks{e-mail \tt{ niall@ucc.ie}}}
\address{
${}^{(1)}$\
 Instituto de Ciencias Nucleares \\
 Universidad Nacional Aut\'onoma de M\'exico\\
 Apdo. Postal 70-543, 04510 M\'exico, D.F., MEXICO \\
${}^{(2)}$\ Physics Department,  University College Cork,\\
Cork, IRELAND \\
}
\maketitle
\begin{abstract}
We exploit an arbitrary extrinsic
time foliation of spacetime to solve the constraints
in spherically symmetric general relativity.
Among such foliations there is a one parameter family,
linear and homogeneous in the extrinsic curvature,
which permit the momentum constraint to be solved exactly.
This family  includes, as special cases, the extrinsic time gauges that
have been exploited in the past. These foliations have the
property that the extrinsic curvature is spacelike with respect to the
the spherically symmetric superspace metric. What is
remarkable is that the linearity can be relaxed at no essential extra cost
which permits us to isolate a large
non - pathological dense subset of all extrinsic time foliations.
We identify properties of solutions which are
independent of the particular foliation within this subset.
When the geometry is regular, we can place spatially invariant numerical 
bounds on the values of both the spatial and the temporal gradients
of the scalar areal radius, $R$. These
bounds are entirely independent of the particular gauge
and of the magnitude of the sources.
When singularities occur, we demonstrate
that the geometry behaves in a universal way
in the neighborhood of the singularity.
These results can be exploited to develop
necessary and sufficient conditions for the existence of both
apparent horizons and singularities in the initial data which
do not depend sensitively on the foliation.
\end{abstract}
\date{\today}


\section{ INTRODUCTION}

In this paper we examine the constraints in general relativity
when the  spatial geometry is spherically
symmetric and possesses just one
asymptotically flat region \cite{I,II}.
This is the simplest gravitational scenario which exhibits
local degrees of freedom.

In \cite{II} we focused on the solution of the constraints
when the extrinsic curvature $K_{ab}$ vanishes. Though they are simple,
they nonetheless  display some of the features of the general problem.
Such solutions are, however, very special. For if
the extrinsic curvature vanishes, the momentum constraint
requires that the current
density of the matter fields, $J$, must also vanish.
The solutions of the Hamiltonian constraint which result
therefore correspond to `momentarily static' spatial geometries
which do not generally occur in a given spacetime \cite{MOTS}.
Since, if they occur at all, they
occur  as isolated objects, we did not
need to fix the foliation.
Here, we extend our work to cover the general situation where matter
flows and, as a result, the extrinsic curvature is non-vanishing.
The advantage of having
dealt separately with the momentarily static  solutions is
that we can focus here on the physical feedback on the spatial
geometry introduced by extrinsic curvature.

The introduction of extrinsic curvature complicates the analysis
substantially. This occurs on two levels.
The first is purely technical: the
Hamiltonian constraint gets coupled to the momentum constraint ---
we have to solve a coupled system of equations.
The second is conceptual:
the constraints do not single out a unique
slice through the spacetime --- we
need to specify some foliation.

In general, the initial data is given by
specifying the intrinsic and the extrinsic geometry on some
spacelike hypersurface which satisfy the constraints.
A spherically symmetric geometry is completely characterized by
specifying the areal radius $R$ as a function of the
proper radius, $\ell$.
The extrinsic curvature can be expressed in a form
consistent with spherical symmetry

\begin{equation}
K_{ab}= n_a n_bK_{\cal L} + (g_{ab}-n_a n_b)K_R\,,
\label{eq:scalars}\end{equation}
where $K_{\cal L}$ and $K_R$ are two spatial scalars and
$n^a$ is the outward pointing unit normal to the two-sphere of fixed
radius in the slice.

How does one go about fixing the foliation?
In principle, any foliation admitting globally regular solutions of the
constraints is as good as any other.
Ideally, therefore,  we would like to
consider a completely general slice through the spacetime;
realistically, however, if the slicing is too general, it
becomes very difficult to prove anything.
At very least, the feedback on the spatial geometry introduced by extrinsic
curvature should reflect the strength of the material
currents flowing ---  the gauge certainly should not
overshadow completely the underlying physics.

We will focus in this
paper on an `extrinsic time' foliation
of spacetime. This involves fixing some spatial scalar function of
the extrinsic curvature. For a a spherically symmetric geometry, we
can cast this relationship in terms of
two scalars appearing in (\ref{eq:scalars}) as follows

\begin{equation}
{\cal F} [K_R, K_{\cal L}] = 0\,,
\label{eq:fofK}\end{equation}
with a possible dependence on $R$ and $\ell$
which we have not indicated explicitly.
Any gauge of this form
should (at least implicitly) be solved to fix one of
the scalars ($K_{\cal L}$ say)
appearing in (\ref{eq:scalars}),
in terms of the other.

All previous work on the constraints in
spherically symmetric relativity has focused exclusively on some given
foliation of this type. These have been maximal slicing or the
so called polar slicing \cite{refI}.
The latter slicing mimics the $K_{ab}=0$
form of the Hamiltonian constraint and is the foliation which
provides the standard presentation of the Schwarzschild
geometry. Unfortunately, in either case, one is at a loss to know
just how sensitively the solution  depends on the choice of gauge.
How will our notions of the  size and the energy content
change in another foliation? If they change
in a way we cannot quantify they are almost useless.
To address this kind of question
it is desirable to work with as large
a class of foliations as possible.

In \cite{I}, we introduced a function $\alpha$, the ratio
of the two scalars defining the extrinsic curvature

\begin{equation}K_{\cal L} + \alpha\, K_R = 0. \label{eq:alpha}\end{equation}
By setting $\alpha$ equal to some specified function,
$\alpha=\alpha[K_R, R, \ell]$ say,
Eq.(\ref{eq:alpha}) defines an extrinsic time foliation.
If $\alpha$ is a function only of $R$ and $\ell$
(in particular, if it is constant),
the momentum constraint is exactly solvable for
$K_R$.

It was shown in \cite{I} that each
constant value of $\alpha$ which is greater than $0.5$
provides a globally regular slice
for appropriate sources. These values of $\alpha$
correspond to a spacelike extrinsic curvature
`vector' with respect to the superspace metric.
As special cases we recover both the maximal slicing,
with $\alpha = 2$, and the polar gauge
when $\alpha \to \infty$.

Remarkably, one can show that even when $\alpha$
is not a constant the gauge continues to provide
regular slices of spacetime so long as $\alpha\le 0.5$
asymptotically. All spacelike extrinsic curvatures in superspace
provide regular foliations.
The identification of potentially singular geometries
will, however, require that the gradients of $\alpha$ be appropriately
bounded.

The gauges we consider,
in fact, represent a very large class of extrinsic time foliations.
Recasting (\ref{eq:fofK})
in the homogeneous form,
$K_{\cal L} = -\alpha [K_R, R, \ell]\, K_R$,
ensures that when the material current $J$ vanishes, $K_{ab}=0$.
In particular, flat spacetime will be foliated by flat spatial hypersurfaces.
Indeed, when the momentum constraint
is satisfied, the extrinsic
curvature is (quasi-) linear in $J$,
albeit in a non-local way.
In this way the extrinsic curvature of the
hypersurface responds directly to the movement of matter on it ---
a physically reasonable criterion.

Within this large class of extrinsic time foliations,
there are universal properties exhibited by
solutions of the constraints
which are either independent of, or do not depend
sensitively on the particular foliation.
These properties  divide naturally into
those of globally regular geometries and those of singular geometries.

In \cite{II}, we examined these properties
when the initial data was momentarily static.
We first identified a geometrical bound on the
spatial gradient of the areal radius $R$ (the prime is a derivative with
respect to proper radius),

\begin{equation}
-1\le R' \le 1\,,\label{eq:R'1}\end{equation}
independent of the source
which was valid in all globally regular geometries.
This bound was seen to operate at a more fundamental level
than the positivity of the ADM mass.
We then went on to investigate how
the matter content of the slice can potentially force
the appearance of either apparent
horizons or singularities.
We showed that when singularities occurred,
they possessed a universal form and we could place bounds on the rate of
divergence of geometrical scalars.

How do these results generalize?

In globally regular geometries, the spacetime
gradients of $R$ are bounded.
Firstly, when the weak energy condition is satisfied and $\alpha\ge 0.5$
everywhere, the bound Eq.(\ref{eq:R'1}) on the spatial gradient
continues to hold. Secondly, and
perhaps more surprising, an analogous bound can be
placed on the extrinsic curvature
when the dominant energy condition is satisfied.
We obtain the highly non-trivial result that

\begin{equation}
-1\le R K_R \le 1\,,\label{eq:KR1}
\end{equation}
if $\alpha \ge  1$.
The bound  (\ref{eq:KR1}) can be interpreted as a bound on
$\dot R$, the derivative of $R$  with respect to normal proper timeBoth of
these bounds are independent of the source magnitude.
They will play a central role in the establishment of
sufficient conditions for the appearance of apparent horizons
and singularities \cite{IV}.
If $\alpha<1$, no such bound exists
--- indeed, a counterexample can be constructed.

Singular geometries can occur even though both $\rho$ and $J$ are finite.
The only way that the geometry can become singular, however,  is by
pinching off at some finite proper radius from the center.
Generically, at this radius ($\ell_S$ say), $R$ will vanish non-analyically

\begin{equation}R\sim
C (\ell_S-\ell)^{1\over\alpha+1}
\,.\label{eq:RS}\end{equation}
Remarkably, the
quasi-local mass (QLM) remains finite even when
the geometry is singular. Indeed, we show that this is
always true regardless of the gauge condition.
Our ability to identify universal behavior of this
form will be crucial for the establishment of necessary conditions for
singular geometries in a subsequent publication \cite{V}.

Generally, the singularities of the
three-geometry consistent with the constraints will be more severe
than those which are admissable at a moment of time symmetry.
If, however, the movement of matter is tuned so that the extrinsic
curvature vanishes as the singularity is approached,
the strength of the singularity will be determined entirely by the
QLM, exactly as it
is at a moment of time symmetry \cite{II}. We
show that this tuning
corresponds to an integrability condition on the current.
If, in addition, the tuning
is refined so that the QLM also vanishes as we approach the
singularity the
curvature singularity disappears and the spatial geometry
pinches off in a regular way. This latter integrability condition
involving the QLM is completely analogous to the integrability
condition encountered at a moment of time symmetry.
Regularity at the singularity is, of course, precisely the condition
that the interior be a regular closed universe. If the matter fields
carry conserved charges these will, in their turn, have integrability
conditions associated with them. Viewed this way, regular closed universes
appear to be very special universes \cite{GOM}.

The paper is organized as follows:

We begin in Sect.2 with a discussion of
the solution of the momentum constraint.
In Sect.3 we provide a derivation of the
bounds on $R'$ and $RK_R$.
In Sect.4, we derive Eq.(\ref{eq:RS}).
In Sect.5, we discuss the integrability conditions
and comment briefly
on the regularity of Euclidean relativity.
We conclude in Sect.6  with a brief discussion and
outline of future work.

\section{ THE CONSTRAINTS}

In this section we examine the analytical structure
of the constraints when $K_{ab}\ne 0$.

We recall that the constraints can be written as

\begin{equation}
K_R\left[K_R+2K_{\cal L}\right]-
{1\over R^2}\Big[2 \left(R R^\prime \right)^\prime -R^{\prime 2}
-1 \Big]=8\pi \rho\label{eq:hamilton}\end{equation}
and
\begin{equation}K_R^\prime +
{R^\prime \over R}(K_R-K_{\cal L})=4\pi J\,,\label{eq:mom}\end{equation}
where the line element on the spatial geometry is parametrized by
\begin{equation}ds^2= d\ell^2+R^2 d\Omega^2\,,\label{eq:lineel}\end{equation}
$\ell$ is the proper radial distance on the hypersurface,
$R$ is the areal radius,
and we have expanded the  extrinsic curvature
according to Eq.(\ref{eq:scalars}).
All derivatives are with respect to the proper radius of the
spherical geometry, $\ell$.  The spatial geometries we
consider consist of a single asymptotically flat
region with a regular center, $\ell=0$. We will subsequently
refer to such geometries as regular. The appropriate boundary
condition on the metric at $\ell=0$ is then
\begin{equation}R(0)=0\,.\label{eq:bc}\end{equation}
We recall that $R^\prime(0)=1$ if the geometry is regular at this
point.
We assume that both $\rho$ and $J$ are appropriately bounded
functions of $\ell$ on some compact support. This  compact support
restriction could be easily relaxed so as to consider solutions where
both $\rho$ and $J$ decay appropriately as one approaches infinity with
little extra effort but also with little extra insight.

We define
the quasi-local mass $m$ as follows

\begin{equation}m = {R\over 2} \Big(1- R^{\prime2}\Big) + {1\over 2} K_R^2 R^3
\,.
\label{eq:quasiloc}\end{equation}
When the constraints Eq.(\ref{eq:hamilton}) and
(\ref{eq:mom})
and the boundary condition at the origin, Eq.(\ref{eq:bc})
are satisfied, $m$ is determined by the sources as follows:

\begin{equation}m=4\pi \int_0^{\ell} d\ell\, R^2\Big[\rho R^\prime  + J R K_R
\Big]
\,.\label{eq:intform}\end{equation}
This way, $m$ arises as a first integral of the constraints.
These equations are gauge invariant.
In a globally
regular geometry, $m$ coincides at infinity with the ADM mass, $m_\infty$.
As we found in \cite{II} in a simpler context, the
introduction of $m$ is extremely useful and will be exploited
repeatedly in our analysis.

To solve the constraints classically, we need to specify some
foliation. In this paper, we will
focus on a gauge condition of
the general form (\ref{eq:alpha}) where $\alpha$ is some
specified function of the configuration variables,
$R$, $K_R$ and $\ell$.
It is possible to provide a geometrical interpretation for these gauges.
To begin with, we know that when $\alpha=2$ this condition
specifies a spacelike  hypersurface with maximum volume in spacetime:
the trace of the extrinsic curvature $K= K_{\cal L} + 2K_R$ vanishes.
It is simple to show that when  $\alpha =\alpha(\ell)$,
Eq.(\ref{eq:alpha})  is precisely the condition that the
modified spatial volume of a closed ball,

\begin{equation}
V_\alpha = 4\pi\int_0^\ell d\ell\, R^\alpha \,,
\end{equation}
be a maximum.

When $\alpha$ is a constant,
the momentum constraint can be solved uniquely for $K_R$ in
terms of the radial flow of matter, $J$, as follows

\begin{equation} K_R =
{4\pi\over R^{1+\alpha}}\int_0^{\ell} d\ell\, R^{1+\alpha} J
\,,\label{eq:mom1}\end{equation}
where we have exploited the regularity of the
geometry at the origin to set $K_R(0)=0$.

When Eqs.(\ref{eq:alpha})  and (\ref{eq:mom1}) are substituted into
Eq.(\ref{eq:hamilton}),
we obtain a second order singular non-linear integro-ODE for
$R$ \cite{order3}.
Subject to the boundary condition, (\ref{eq:bc}), the solution is
uniquely determined. Not only is the extrinsic curvature
completely determined by the material sources, so also is the
spatial geometry. There are no independent gravitational
degrees of freedom, exactly as expected.
We note that in the gauge Eq.(\ref{eq:alpha}), the spatial geometry
does not depend on the global sign of $J$. Of course, if we reverse
the sign of $J$, the extrinsic curvature picks up a negative sign.

When $\alpha$ is not  constant,  it is still possible to
mimic the solution when $\alpha$ is constant.
To do this we recast the momentum constraint in the
form

\begin{equation}
(R^{1+\alpha}K_R)' = 4 \pi
R^{1+\alpha}J   + \alpha'\ln (R/L) R^{1+\alpha} K_R\,.
\end{equation}
where $L$ is any characteristic  length scale. The
spatial variation of $\alpha$ has been absorbed completely
into the second term on the RHS. The solution is given by

\begin{equation}
K_R = {4\pi\over R^{1+\alpha}}\int_0^\ell d\ell_1\,
R^{1+\alpha} J \,\Delta (\ell_1,\ell)
\,,
\label{eq:mom2}\end{equation}
where

\begin{equation}
\Delta (\ell_1,\ell)= e^{\int_{\ell_1}^\ell d\ell_2 \,\alpha'\ln (R/L)}
\,.
\label{eq:Delta}
\end{equation}
The constant $\alpha$ result is simply modulated
by an exponential multiplicative correction, $\Delta$.
We note in particular that $K_R=0$ when $J=0$.
This provides a very strong justification for
casting the gauge in the form, (\ref{eq:alpha}).
If $\alpha=\alpha(R,\ell)$ alone, $K_R$ will
also be linear and homogeneous in
$J$. If we admit a $K_R$ dependence explicitly into $\alpha$
the linear correlation of $K_R$ and $J$ no longer holds.
$K_R$ will nonetheless be positive when $J$ is.
If, however, $\alpha$ is an even function of
$K_R$ then $K_R$ will echo the parity of
$J$: $K_R[-J]= -K_R[J]$.

It is clear from inspection of the definition
of $\Delta$ that spatial variations in $\alpha$ are
anti-screened: remote source contributions get distorted
more than nearby ones. This is potentially worrysome
but, as we will see, it is not a serious obstacle.

We will now look more closely at the analytic structure of
solutions.

Let us first focus on regular geometries.

\section{ GLOBALLY REGULAR GEOMETRIES}

We first comment on the behavior of the spatial geometry
in the neighborhood of the origin.
In fact, in the neighborhood of
$\ell=0$, we have $R\sim \ell$, so that \cite{J}

\begin{equation}K_R\sim 4\pi {J(0)\over 2
+\alpha(0)} \ell\,.
\label{eq:K0}\end{equation}
Thus if $K_R$ is regular at the origin
then it must also vanish there. This is not, however,
surprising. Spherical symmetry is very restrictive leaving
a regular geometry no freedom to evolve at the origin.

We can expand $R(\ell)$ in a power series
in the neighbourhood of $\ell = 0$ and substitute into
Eq.(\ref{eq:hamilton}) to get

\begin{equation}R(\ell) = \ell + {4\pi \over 9} \rho(0) \ell^3
+ \dots \label{eq:R0}\,.\end{equation}
A consequence of the vanishing of $K_R(0)$
is that $J$ will only show up at order five in this expansion
--- two orders behind $\rho$. The  metric at the origin
clearly is not sensitively dependent on the current flowing there.

The other region we need to check is outside the source.
What constraints does asymptotic flatness place on $\alpha$?
To recover an asymptotically flat spatial geometry we require
that $R(\ell)\sim\ell$ to leading order.
We have that $K_R\sim {\rm constant}\,\Delta(\ell_0,\ell)/R^{1+\alpha}$.
Now for an appropriate falloff
(faster than $\ell^{-1}$) on $\alpha'$, $\Delta(\ell_0,\ell)$
will always saturate so that we can absorb it into the constant.
On one hand, we note that
outside the source, the integral identity (\ref{eq:intform})
implies  $m$ is a constant. However, the contribution of extrinsic
curvature to $m$ (Eq.(\ref{eq:quasiloc}))
$\sim 1 /R^{2\alpha_\infty - 1}$.
There represents an inconsistency
if $\alpha$ tends asymptotically to any value, $\alpha_\infty$,
lower than $0.5$.
If $\alpha_\infty>0.5$, not only $K_R$ but also its contribution to $m$
vanish asymptotically.
Such a choice is simultaneously regular at the origin.

We note that  with strict inequality, $\alpha_\infty > 0.5$,
$m$ be dominated asymptotically by the
first term in (\ref{eq:quasiloc})
so that $m_\infty$ is encoded completely in the intrinsic geometry.
In the limiting case,
$\alpha=0.5$, the intrinsic and the extrinsic geometries
share the burden. However, such a falloff
invalidates the traditional expressions for the ADM mass.

What can we say in general about globally regular geometries?

We will demonstrate that they possess the remarkable property
that for an appropriate dense subset of extrinsic time foliations
both the spatial
and temporal gradients of $R$ are bounded numerically
in a way which is entirely independent of the material sources and of $\alpha$.

Suppose $\rho$ satisfies the weak energy condition, $\rho\ge 0$.
Consider any foliation satisfying Eq.(\ref{eq:alpha})
with $\alpha \ge 0.5$ everywhere.  Then, if the geometry is regular,
$R'^2\le 1$ everywhere.
The  proof is very simple and was given in I. We
repeat it here to emphasise that the spatial variation of
$\alpha$ does not enter:
$R'$ must be bounded in any
regular geometry.
We note that $R'(0)=1$ and $R'\to 1$ at infinity. Thus
$R'$ must possess an interior critical point. At this point
$R''=0$.  In the gauge (\ref{eq:alpha}),
the Hamiltonian constraint, Eq.(\ref{eq:hamilton}) now reads
at this point

\begin{equation}R'^2
 = 1 - 8\pi\rho R^2
- (2\alpha -1) R^2 K_R^2\,.\label{eq:R'max}\end{equation}
Both the second and the third terms on the RHS are
negative. The result follows immediately.

A simple corollary of this result is the positivity of $m$.

We can obtain analogous bounds on $K_R$ in the gauge Eq.(\ref{eq:alpha}).
However, it is important to stress that without some control over
$J$ we should not expect $K_R$ to be bounded.
Let us therefore suppose that the dominant energy condition (DEC),
$|J|\le \rho$ is satisfied everywhere.
Our experience in \cite{I} suggests that
when the DEC holds, the appropriate
variables are the optical scalars defined by \cite{MOM},

\begin{equation}
\omega_\pm =  2\Big(R^\prime \pm  R K_R\Big)\,,\label{eq:omegapm}
\end{equation}
which are ($R$ times) the null expansions in the out-future and
out-past directions.
They are a useful set of variables to exploit when we are interested
in identifying apparent horizons \cite{I,MOM,IV}.
The optical scalar which
marks the presence of a future trapped surface is $\omega_+$:
$\omega_+ = 0$ at a future apparent horizon.

It was shown in \cite{MOM} (and again in \cite{I}) that
when the dominant energy condition, $|J|\le \rho$,
is satisfied

\begin{equation}|\omega_\pm|\le
\kappa + \sqrt{|\kappa|^2 +4}\,,
\label{eq:omegamax}\end{equation}
where $\kappa = {\rm Max} |RK|$ and $K$ is the trace of the
extrinsic curvature. These bounds depend on the sources only through $K$.
When $K=0$ ($\alpha=2$) they become numerical
bounds which are independent of the sources,

\begin{equation}
-2\le \omega_\pm \le 2\,.\label{eq:omega2}
\end{equation}
It then follows that

\begin{equation}
R K_R = (\omega_+ - \omega_-)/4\,,
\label{eq:Komega}\end{equation}
and $R' = (\omega_++\omega_-)/4$ are
bounded, $|R K_R | \le 1$ and $|R'|\le 1$.

Here, we would like to show that
even when $\alpha\ne 2$ in the gauge Eq.(\ref{eq:alpha})
it is still possible to bound $\omega_\pm$ by
Eq.(\ref{eq:omega2}).
One way to do this is to bootstrap on Eq.(\ref{eq:omegamax}).
This way, one can bound $|\omega_\pm|$ when
$\alpha$ lies within the range  $1 <\alpha < 3$. However, the
bound does depend on $\alpha$ and diverges at the points,
$\alpha=1$ and $\alpha=3$.
However, one can do better.

We showed in \cite{I}
that it is possible to add and subtract
the two constraints (\ref{eq:hamilton}) and
(\ref{eq:mom}) to obtain simple equations for the
spatial derivative of $\omega_+$ and $\omega_-$.

\begin{equation}
(\omega_\pm)^\prime=  -8\pi R(\rho \mp J) -{1\over 4R}
\left(\omega_+\omega_- - 4 \right) \pm
\omega_+ K_{\cal L}\,,\label{eq:constraints}\end{equation}
We now exploit the gauge condition,
(\ref{eq:alpha}) and (\ref{eq:Komega}) to recast these
equations in the form,

\begin{equation}
(\omega_\pm)^\prime=  -8\pi R(\rho \mp J) + {1\over 4R}
\left((\alpha-1)\omega_+\omega_- + 4 - \alpha \omega_\pm^2\right)
\,.
\label{eq:constraints1}\end{equation}
Let us establish the inequality (\ref{eq:omega2}) for $\omega_+$.
The argument is very similar to the one we used above to
derive the bound $R'^2\le 1$. We note that
$\omega_+$ must be bounded in a regular geometry and that
$\omega_+(0)=2$ and $\omega_+\to 2$ at infinity. Thus
$\omega_+$ must possess an interior critial point. At this point
$\omega_+'=0$ so that

\begin{equation}
(\alpha-1)\omega_+\omega_- + 4 - \alpha \omega_+^2
=32\pi R^2(\rho-J)
\end{equation}
the right hand side of which is positive by hypothesis. Thus

\begin{equation}
\alpha \omega_+^2
\le
(\alpha-1)\omega_+\omega_- + 4 \,.\label{eq:i}
\end{equation}
However, the quasilocal mass is positive, or equivalently \cite{I,MOM}

\begin{equation}
\omega_+\omega_- \le 4\,,
\label{eq:4}
\end{equation}
so that when $\alpha\ge 1$, $\omega_+$ satisfies
Eq.(\ref{eq:omega2}).

In \cite{I}, we pointed out that the `binding energy'
$M-m$ of a regular spherically symetric system satisfying
the dominant energy condition is positive when the slicing is
maximal. In general, we have

\begin{equation}
M-m = 4\pi \int_0^\ell d\ell\,
R^2\left[(\rho+ J)(2-\omega_+) +
         (\rho-J) (2+\omega_-)\right]\,,
\end{equation}
which is manifestly positive whenever $|\omega_\pm|\le 2$.
Thus this result is also true for all values of $\alpha\ge 1$.

It is clear from Eq.(\ref{eq:i}) that Eq.(\ref{eq:omega2})
cannot be extended to $\alpha<1$ ---
$\omega_-\omega_+$ is not bounded from
below. To obtain a bound we need to exclude both future and past trapped
surfaces so that $\omega_+\omega_-$ is positive.
We then have for all positive $\alpha$,

\begin{equation}
|\omega_+| \le 2{\rm Max}\Big(2/\sqrt{\alpha}_{\rm Min},1\Big)
\,.\label{eq:alp<1}
\end{equation}
The results for $\omega_-$ is identical.

This implies the bound on $K_R$:

\begin{equation}
|RK_R|\le {\rm Max}\Big(1/\sqrt{\alpha_{\rm Min}},1\Big)\,.
\end{equation}
We already have $R'^2\le 1$ for all $\alpha$ when the weak energy condition is
satisfied.

One can, in fact, easily construct a counterexample
demonstrating explicitly that we should not expect to do better
than Eq.(\ref{eq:alp<1}) when
$\alpha<1$. We do this by examining the values
assumed by the optical scalars in the neighborhood of $\ell=0$.
We can combine Eq.(\ref{eq:K0}) and (\ref{eq:R0})
to obtain

\begin{equation}\omega_\pm \sim 2 - {8\pi\over 3}(\rho (0) \mp
{3\over 2+\alpha}J(0) ) \ell^2 \,.\label{eq:2.8}\end{equation}
If the dominant energy condition is satisfied, then when
$\alpha\ge 1$, we have  $\omega_\pm \le 2$ near the origin
which is consistent with our result.
If, however, $\alpha<1$ this is not necessarily the case. If
$J(0)$ exceeds $(2+\alpha)\rho(0)/3$ and $\alpha <1$,
then $\omega_+\ge 2$ in the neighborhood of the origin.
We note
also that

\begin{equation}\omega_+\omega_- \le (2 - {8\pi\over 3}[\rho (0)\ell^2])^2
- \left({8\over 2+\alpha}\right)^2
\pi^2 J(0)^2 \ell^4 \le 4 \,.\end{equation}
This is consistent with the inequality
Eq.(\ref{eq:4}).
Note also that the absolute maximum of the
product $\omega_+\omega_-$ obtains at the boundary values $\ell=0$ and
$\ell=\infty$ and it is also the flat space value. When $K=0$,
this is also true of
both $\omega_+$ and $\omega_-$. In general, the absolute maximum of
neither need occur at these points.

\section{ SINGULAR GEOMETRIES}

So far we have assumed that the geometry is regular everywhere.
A non-singular asymptotically flat solution defined for all
$\ell\ge 0$ will not, however, always exist for every specification of $\rho$
and $J$. In this section, our task will be to understand what can go wrong
and to identify the mechanism driving the geometry into a singularity.

In Sect.3 we showed that $R'^2\le 1$ in any globally regular geometry.
Thus if $R^{\prime2}>1$ anywhere the geometry must be singular.

Let us suppose that $R^{\prime2}>1$ at some point.
Then,  when $K_{ab}$ satisfies
Eq.(\ref{eq:alpha}), Eq.(\ref{eq:hamilton}) implies

\begin{equation}
R R'' =
{1\over 2}\Big[1-R^{\prime 2}
\Big] + {R^2\over 2} (1-2\alpha) K_R^2
- 4\pi \rho\,,\end{equation}
so that $R^{\prime\prime} <0$ and
$R^\prime$ is decreasing there. This can only occur
by $R^\prime$ falling through $R^\prime=-1$. Once $R^\prime$
falls below this value it will continue decreasing
monotonically thereafter. The surface with $R^\prime=-1$ in the
configuration space therefore acts as a oneway membrane.
Suppose that the areal radius is $R_0$
when  $R^\prime=-1$. We know now that the solution must
crash, {\it i.e.} $R\to 0$ within a finite proper distance
which is less than or equal to $R_0$ from that point.

Since $R'' \le 0$ whenever $R' = 1$ we see that the surface $R' = 1$
in the configuration space also acts as
a oneway membrane and the solution can only
pass downwards through it. However, since at a regular center we have $R' = 1$
and $R'$ starts to reduce as soon as we enter matter, it is clear that the
region defined by $R' > 1$ is completely forbidden.

We conclude that  crashing through
$R'=-1$ is the  generic way the spatial geometry can become
singular. Singularities with $R^\prime=-1$ at $R=0$ are also possible.
They result, however, only for special finely tuned
matter distributions. We will discuss them more fully below in Sect.5.

Putting the regular and singular results together, we have the following:
if the geometry is globally regular, then $-1 \le R'\le 1$ everywhere;
if $-1< R^\prime \le 1$ everywhere, then the
geometry is globally regular.

One possible way that this method of constructing initial
data for a spherically symmetric gravitational field can
break down is that the slicing turns null. Since $R$ is a
four-dimensional scalar, nothing will go bad with it.
On the other hand,
$\ell$ is the spacelike proper distance along the slice and so $d\ell$
will become small and thus $R'$ will become unboundedly large if the slice
turns null. But we have shown that this cannot happen if we assume
$\alpha \ge 0.5$
and $\rho \ge 0$. Note that we do not have to assume that $\alpha =$
constant. Therefore, if we stay inside the lightcone of the
super-metric, we stay outside the lightcone
of the spacetime! For any slice satisfying $\alpha \ge 0.5$
the only possible singularity is when $R \rightarrow 0$.

Let us now examine more carefully the approach towards a
singularity.
In the neighborhood of the point $\ell=\ell_S>0$ at which
$R=0$, Eq.(\ref{eq:mom2}) implies that

\begin{equation}K_R\sim
{C_\alpha(\ell_S)\over R^{1+\alpha}}\,,\label{eq:2.13}\end{equation}
where
\begin{equation}C_\alpha(\ell)= 4\pi\int_0^{\ell} d\ell_1\, JR^{1+\alpha}
\Delta(\ell,\ell_1)
\label{eq:C}\end{equation}
is finite if $\Delta$ is. $K_R$ will therefore
be singular (for physically acceptable values of $\alpha$)
if the geometry pinches off unless the current is tuned such that

\begin{equation}C_\alpha(\ell_S) =0\,.\label{eq:C=0}\end{equation}

To examine the structure of singularities it is
extremely useful to exploit the definition of the quasi-local mass
introduced earlier.
>From a functional point of view, Eq.(\ref{eq:quasiloc}) is identical to the
energy integral in classical mechanics.
To exploit this analogy, we therefore recast this equation
as follows:

\begin{equation}R^{\prime2}=1-{2m\over R} + K_R^2 R^2\,,
\label{eq:quasiloc'}\end{equation}
where $m$ is given by Eq.(\ref{eq:intform}) and $K_R$ by Eq.(\ref{eq:mom1}).

Let us suppose that $m$ remains finite.
Now, if $C_\alpha$ does not vanish
and $\alpha>0.5$, the most
singular
term in Eq.(\ref{eq:quasiloc'}) is the quadratic in $K_R$. This implies that

\begin{equation}R^{\prime2}\sim R^2 K_R^2\label{eq:R'atsing}\end{equation}
in the neighborhood of $R=0$, or
$R^{\prime2} \sim C_\alpha^2/  R^{2\alpha}$.
Generically, therefore, $R^{'2}$ diverges.
The solution is

\begin{equation}R\sim
\left({ C_\alpha\over\alpha+1}\right)^{1\over\alpha+1}
(\ell_S-\ell)^{1\over\alpha+1}
\,.\label{eq:Ratsing}\end{equation}
If $\alpha>0.5$, such spatial singularities are more
severe than the strong singularities discussed in \cite{II}
which are consistent with
the Hamiltonian constraint at a moment of time symmetry.
We will refer to the generic kind of singularity
driven by extrinsic curvature as a strong $J$-type singularity.
As $\alpha$ increases, the power law determining the strength of
the singularity increases.
Note that the limit $\alpha\to\infty$ (the polar gauge discussed
in I) is extremely singular. This is, however, a gauge artifact
reflecting how poor the polar gauge really is.
Unlike the strong singularities occurring when $K_{ab}=0$,
at which the
scalar curvature ${\cal R}$ remained finite, ${\cal R}$ will generally
blow up (just like $ K_R^2 \sim 1/(\ell_S-\ell)^2$). On dimensional
grounds, we expect
all curvature scalars to blow up as $1/(\ell_S-\ell)^2$ as we
approach a singularity unless there is some constraint
obstructing them from doing so.

To show that the above analysis is self- consistent,
 we need to demonstrate that:
for finite $\alpha'$, (i)the form factor $\Delta$ defined by
Eq.(\ref{eq:Delta}) remains finite; (ii)  $m$ remains
finite.

To this end, we  note that

\begin{equation}
\int_0^\ell d\ell\,\alpha' \ln R/L \le |\alpha'|_{\rm Max}
\int_0^\ell d\ell \,|\ln R/L|\,.
\label{eq:Delta1}\end{equation}
The integrated logarithm is bounded. While the integrand diverges
at $R=0$, the integral is nevertheless well behaved. We note that

\begin{equation}
\int_0^s ds\, \ln s = s\,\ln s -s\,.
\end{equation}
Thus in particular, the form factor is well behaved on the flat
solution, $R=\ell$.
This is essentially all we need to check
because as we have just seen $R(\ell)
\sim (\ell_0 - \ell)^{1/1+\alpha_0}$
at a singularity so that
logarithm is multiplicatively identical to the
flat space value.

We now confirm that
$m$ remains finite as we approach a strong
singularity. We do this by demonstrating that the volume integral
(\ref{eq:intform}) is
always finite. We note that for suitably bounded $\rho$ and $J$,

\begin{equation}
(\rho R^2 R^\prime, J R^3 K_R)\sim
(\rho,J) (\ell_S-\ell)^{-\left({\alpha-2\over \alpha+1}\right)}
\,.\label{eq:matsing}\end{equation}
If $\alpha\le 2$, the integrand itself remains finite.
In general, the integral will be finite if the exponent of
$(\ell_S-\ell)^{-1}$ is bounded by one. But ${(\alpha-2)/(\alpha+1}) < 1$
for all finite values of $\alpha$ thus guaranteeing that the integrals
over $R^2 \rho R^\prime$ and $R^3 J K_R$ converge.
The only possible gap in this argument is the assumption that $m$
remains finite. It is possible that $m$ diverges fast enough when the
singularity is
approached so that $m/R$ dominates $R^2K_R^2$. This would require $m$ to
diverge faster than $R^{1 - 2\alpha}$. The first term in Eq.(\ref{eq:intform})
cannot give a divergent $m$ as $R^2\rho$ obviously remains finite and
$\int R'd\ell =
R$ in also bounded as we approach the singularity. Therefore we need only
consider the $JR^3K_R$ term. This will diverge like $R^{2 - \alpha}$.
Let us
assume that $R \sim (\ell_S - \ell)^{\beta}$ for some $\beta > 0$.
Then,
from Eq.(\ref{eq:intform}), $m$ will, at worst, diverge like $m \sim  (\ell_S -
\ell)^{(2 - \alpha)\beta + 1}$. Now the requirement that the $m$ term
in
Eq.(\ref{eq:quasiloc'}) dominates implies $(2 - \alpha)\beta + 1 < (2\alpha -
1)\beta <
0$. This in turn gives $-(\alpha + 1)\beta > 1$, a negative $\beta$!
Therefore it is clear that $m(\ell_S)$ is always finite.

The
sign of $m$ will, however, depend on the details of the current flow.
This is obvious from the definition Eq.(\ref{eq:quasiloc}).
Even if $R^{\prime2}>1$, a sufficiently large value of $K_R$
can render $m$ positive. In particular, unlike the value of $m$
assumed
at strong $\rho$-singularities when $K_{ab}=0$
(discussed in \cite{II}) which is always negative, the
sign generally can assume either value. Indeed $m$ need never even be
negative in a singular geometry. Though $R^\prime$ decreases
monotonically, $m$ nonetheless can remain positive. There is no conflict
with the positive QLM theorem. In our examination of
momentarily static configurations in
\cite{II}, we found that $m$ is positive everywhere except at the
origin
or in a neighborhood of it if and only if
the geometry is non-singular. This is a consequence of the
coincidence of the converse of the bounded $R^{\prime2}$ lemma
and the converse of the positive QLM theorem
when $K_{ab}=0$. In the general case, when $K_{ab}\neq 0$,
no such coincidence occurs.

\vskip1pc
\noindent{\bf The mass is finite at $R = 0$ independent of the gauge}
\vskip1pc

We have shown above that the quasi-local mass $m$ is finite even if
the spatial
slice is strongly singular with $R \rightarrow 0$ so long as $(\rho, J)$
remain finite. This result was derived on the assumption that the slice was
chosen to satisfy the gauge condition, (\ref{eq:alpha}) with
$\alpha\ge 0.5$. It turns out
that the finiteness of $m$ holds on any slice. To see this we need  only
consider Eq.(\ref{eq:intform}), which is slicing independent, and
Eq.(\ref{eq:quasiloc'}),
which is effectively the definition of $m$.
As we argued above, the first term in Eq.(\ref{eq:intform})
remains finite as we
approach the singularity. We know that $R^2\rho$ is bounded and
$\int d\ell\, R' = R$ also is well behaved. Thus we need only focus on the
$\int d\ell\,
R^2J RK_R$ term.  We know that $R^2 J$ is bounded so we only need to
control the $RK_R$ term. The necessary control is given by
Eq.(\ref{eq:quasiloc'}). If
$R^2K^2_R$ is the dominant term on the right
hand side of Eq.(\ref{eq:quasiloc'}),
we get that $|RK_R| \sim -R'$ so the integral in Eq.(\ref{eq:intform})
is finite as $R
\rightarrow 0$. If $m$ becomes large and negative so that the term
$-2m/R$ dominates, we get  $|RK_R| < -R'$ so again the integral converges.
The only case left to consider is the possibility that $2m/R$ is positive and
diverges at the same rate as $R^2K^2_R$ and some cancellation occurs so that
$R'$ is uncorrelated with $RK_R$. Let us assume $R \sim (\ell_S -
\ell)^{\beta}$ for some $\beta > 0$ and $|RK_R| \sim (\ell_S -
\ell)^{-\gamma}$ for $\gamma > 0$. We then get $m \sim (\ell_S -
\ell)^{\beta - 2 \gamma}$. However, the argument in Eq.(\ref{eq:intform})
goes like
$R^3K_R \sim (\ell_S - \ell)^{2\beta - \gamma}$. Hence we get $m$
diverging
at worst like $m \sim (\ell_S - \ell)^{2\beta - \gamma + 1}$. If this
is selfconsistent, we require $2\beta - \gamma + 1 = \beta - 2 \gamma$.
This implies $\beta + \gamma + 1 = 0$, which makes no sense. Thus such
cancellation cannot take place.

\section{INTEGRABILITY CONDITIONS}

What are the implications of the integrability condition,
Eq.(\ref{eq:C=0})? If Eq.(\ref{eq:C=0}) is satisfied the strong $J$
singularity is moderated to one which is only strong a la $\rho$.
The behavior in the vicinity of the singularity will then be
determined by the $m/R$ term in Eq.(\ref{eq:quasiloc'}) even if the system was
originally `driven' towards the singularity by extrinsic
curvature. If, in addition,
\begin{equation}m(\ell_S)=4\pi\int_0^{\ell_{S}} d\ell\, \left[
\rho R^2 R^\prime + J R^3 K_R\right] = 0
\,,\label{eq:m=0}\end{equation}
the singularity will be a weak one with $R^\prime(\ell_S)=-1$.
We note that $R''=0$ at this point.
The corresponding bag of gold will be a regular closed universe.
These integrability conditions do depend on $\alpha$. If
a given function $J$ satisfies Eq.(\ref{eq:C=0}) with one
function  $\alpha$, generally it will not satisfy that condition
with any other function. There is no spacetime diffeomorphism
invariant statement of the integration. The integrability condition
need not be preserved by the evolution.

If $J$ is positive (or negative)
everywhere, $C_\alpha(\ell)$ defined by Eq.(\ref{eq:C}) cannot vanish.
Thus, if matter is collapsing or exploding everywhere,
all singularities must be strong $J$-type singularities.
This contrasts with the obstruction, $\rho^\prime <0$,
discussed in \cite{II}, {\it prohibiting} the formation of any singularity
when $K_{ab}=0$. In general, we note that on performing an integration
by parts on the first term, $m$ can be rewritten

\begin{equation}m(\ell)={4\pi\over 3}\rho R^3 + 4\pi \int_0^{\ell} d\ell\, R^3
\left[J  K_R - \rho^\prime\right]
\,.\label{eq:mparts}\end{equation}
The first term is manifestly positive. So is the third if
$\rho^\prime\le 0$. If $J$ is positive (negative)
everywhere then so is $m$ in any $\alpha$ - gauge. However,
the third term appearing on the RHS of Eq.(\ref{eq:quasiloc'})
may still pull the
geometry into a singularity if $J$ is sufficiently large.
The peculiarity of momentarily static configurations with
$\rho^\prime <0$ discussed in \cite{II} can clearly be destabilized by
the motion of matter.
All regular closed cosmologies simultaneously satisfy
two integrability conditions, Eqs.(\ref{eq:C=0}) and (\ref{eq:m=0}).
There can be no net flow of material from one pole to the other.
In particular, $J$ must change sign between the poles.
In addition, Eq.(\ref{eq:mparts}) tells us that
\begin{equation} m(\ell_S)=4\pi\int_0^{\ell_{S}} d\ell\, R^3
\left[J K_R - \rho^\prime \right]=0
\,.\label{eq:mparts0}\end{equation}
In particular, $J K_R - \rho^\prime$ must change sign
between the poles. These conditions will be examined in
the closed cosmological context in a subsequent publication [4].

\vskip1pc
\noindent{\bf There are no strong $J$ singularities in the Euclidean
Theory}
\vskip1pc
The singularity structure we have investigated has
one important consequence for Euclidean general relativity.
If the sign of the quadratic term in $K_R$ appearing in Eq.(\ref{eq:quasiloc'})
had  been negative, {\it instead} of facilitating the occurrence of
singularities it would have presented an obstacle to their
occurrance.
Any non-vanishing extrinsic curvature would therefore tend
to stabilize the spatial geometry against singularity formation.
We note that there is
precisely such a sign switch in the Hamiltonian constraint of
Euclidean
general relativity. The Bianchi identities there tell us that the
solutions
of the constraints represent all possible configurations the system
may assume as it is evolved with respect to Euclidean time.
This suggests that gravitational instantons will tend to
be more regular than their Lorenzian counterparts. In fact, the
most singular Euclidean geometries will occur when the geometry
is  momentarily static. In a tunneling Euclidean
four-geometry, such three-geometries correspond to the initial and
final hypersurfaces of the Lorentzian spacetimes between which it
interpolates.
If these hypersurfaces are themselves non-singular, {\it i.e.}
do not involve Planck scale structures, then
Planck Scale physics does not enter the semi-classical description
of tunneling between them. This would appear to validate
the application of the semi-classical approximation.

\section{CONCLUSIONS}

We have identified a dense subset of extrinsic time foliations with
respect to which there exist universal bounds
on certain geometrical invariants. 
When the geometry is regular, we have described how
the spacetime gradients of $R$ in this dense subset
are bounded numerically, independent both of the gauge and of the sources.
Near a singularity, these gradients diverge in a way we
can quantify.

These results can be applied to address the question of identifying
necessary and sufficient conditions for
the presence of apparent horizons and singularities in the initial data
\cite{IV,V} extending the work of \cite{BMM1,BMM2},\cite{II} and \cite{MOM}.
In the analysis of sufficient conditions for the appearance of trapped
surfaces and singularities, first  the moment-of-time-symmetry
case was examined\cite{BMM1},\cite{II} followed by maximal slices \cite{BMM2}
\cite{MOM}. We find that, not only can we extend
this work to constant $\alpha$ but
to the large class of extrinsic time foliations described by
Eq.(\ref{eq:alpha}) for variable $\alpha$ within the range $0.5\le
\alpha<\infty$. We also find that we can provide very powerful
generalizations of the necessary conditions introduced in \cite{II}
for moment of time symmetry initial data to general
initial data.

There are a number of interesting
spherically symmetric problems we intend
to pursue. A very satisfactory representation of
regular closed solutions of the constraints can be given as closed
bounded trjectories in the $(\omega_+, \omega_-)$ plane. In this
representation $R$ plays a subsidiary role. These variables
suggest a novel approach to the canonical quantization of
spherically symmetric general relativity \cite{Can}.
Indeed, constant $\alpha$ foliations can be exploited
to provide a new description of
the Schwarzschild solution \cite{Sch}.

The next stage is the examination of the classical
evolution. Write down the Einstein equations with respect to
the optical scalar variables. Can we cast the
theory in Hamiltonian form? If the value of these
variables in the analysis of the constraints is anything to go by,
one has every reason to expect that they will
throw light on the solution of the dynamical Einstein equations,
both analytically and numerically. Indeed these variables
have recently been exploited to establish
global existence results \cite{Rend}.

\section*{Acknowledgements}

\noindent{We gratefully acknowledge support
from CONACyT Grant 211085-5-0118PE
to JG and Forbairt Grant SC/96/750 to N\'OM}.


\begin{references}
%
\bibitem{I} J. Guven and N. \'O Murchadha,
{\it Phys Rev} {\bf D52} 758 (1995)
An extensive list of references is provided here.
%
\bibitem{II} J. Guven and N. \'O Murchadha,
{\it Phys Rev} {\bf D52} 776 (1995)
%
\bibitem{MOTS} Solutions of the constraints when
$K_{ab}=0$ are also often referred to as
moment of time symmetry configurations.
%
\bibitem{refI} See ref.[1] for relevant references.
%
\bibitem{IV} J. Guven and N. \'O Murchadha, {\it Sufficient
Conditions for Apparent Horizons in Spherically Symmetric
Initial Data} (1997)
%
\bibitem{V} J. Guven and N.. \'O Murchadha, {\it Necessary Conditions
for Apparent Horizons and Singularities in Spherically Symmetric Initial
Data } (1997)
%
\bibitem{GOM} J. Guven and N. \'O Murchadha, unpublished (1995)
%
\bibitem{order3}
It is possible to write this equation
so that when differentiated once it
is transformed into a  {\it local} third order singular ODE.
However, it is not particularly illuminating to cast it this way.
%
\bibitem{J}
The constraint equations place no restriction on the
behaviour of $J$ at the origin. They are not
disturbed by $J(0) \ne 0$. However, the
continuity equation $d\rho/dt +div J = 0$ may force the time derivative of the
density to diverge and a conical singularity to emerge on evolving this data.
%
\bibitem{MOM}
E. Malec and N. \'O Murchadha
{\it Phys. Rev.} {\bf D50} R6033 (1994)
%
\bibitem{Sch} J. Guven and N. \'O Murchadha, In preparation (1997)
%
\bibitem{BMM1}
P. Bizo\'n, E. Malec and N. \'O Murchadha {\it Phys. Rev. Lett.}
{\bf 61}, 1147 (1988);
{\it Class Quantum Grav} {\bf 6}, 961 (1989);
%
\bibitem{BMM2}
P. Bizo\'n, E. Malec and N. \'O Murchadha {\it Class Quantum  Grav}
{\bf 7}, 1953(1990)
%
\bibitem{Can} The canonical quantization of vacuum spherically
symmetric general relativity with Schwarzschild topology has been
examined by K. Kuchar, {\it Phys. Rev} {\bf D50} 3961 (1994)
%
\bibitem{Sch} J. Guven and N. \'O Murchadha, in preparation (1997)
%
\bibitem{Rend} A. Rendall, {\it Class Quantum Grav} {\bf 12}, 1517 (1995); G. Burnett and A. Rendall, {\it Class Quantum Grav} {\bf 13}, 111 (1996)
%
\end{references}
\end{document}